\documentclass[conference]{IEEEtran}
\IEEEoverridecommandlockouts
\usepackage{cite}
\usepackage{amsmath,amssymb,amsfonts}
\usepackage{graphicx}
\usepackage{textcomp}
\usepackage{xcolor}
\usepackage{multirow}
\def\BibTeX{{\rm B\kern-.05em{\sc i\kern-.025em b}\kern-.08em
    T\kern-.1667em\lower.7ex\hbox{E}\kern-.125emX}}
\usepackage{verbatim}

\usepackage{mathrsfs}
\usepackage{amsthm}
\usepackage{diagbox}
\usepackage{verbatim}
\usepackage{dsfont}
\usepackage{bm}
\usepackage{subeqnarray}
\usepackage{cases}

\usepackage{caption}

\newtheorem{theorem}{Theorem}
\newtheorem{lemma}{Lemma}
\newtheorem{corollary}{Corollary}

\theoremstyle{definition}
\newtheorem{remark}{Remark}
\newtheorem{definition}{Definition}
\newtheorem{example}{Example}
\newtheorem{proposition}{Proposition}

\usepackage{algorithm}
\usepackage{algorithmic}
\usepackage{bbm}
\usepackage{hyperref}
\hypersetup{colorlinks = true
            linkcolor=red,
            anchorcolor=blue,
            citecolor=green}
              
\newcommand{\Fqn}{\mathbb{F}_q^n}
\newcommand{\Ftn}{\mathbb{F}_2^n}
\newcommand{\Cp}{C^{\perp}}
\newcommand{\F}{\mathbb{F}}
\newcommand{\GC}{\mathcal{G}_C}
\newcommand{\GCp}{\mathcal{G}_{C^{\perp}}}
\newcommand{\GCr}{\mathcal{G}_{C,r}}
\newcommand{\wt}{\operatorname{wt}}

\newcommand{\tr}{\operatorname{tr}}
\newcommand{\supp}{\operatorname{supp}}

\newcommand{\FK}{\operatorname{FK}}

\usepackage[commandnameprefix=ifneeded]{changes}

\hypersetup{hidelinks}

\begin{document}

\title{Graph-Theoretic Approaches to Coding Bounds via Coset Graphs}

\author{
   \IEEEauthorblockN{Zicheng Ye\IEEEauthorrefmark{1}\IEEEauthorrefmark{2}}
  \IEEEauthorblockA{\IEEEauthorrefmark{1}
                     School of Mathematical Sciences, University of Chinese Academy of Sciences}
   \IEEEauthorblockA{\IEEEauthorrefmark{2}
                     Academy of Mathematics and Systems Science, CAS }
                     
    Email: yezicheng@amss.ac.cn
      
\thanks{This work was supported by the National Key R\&D Program of China (2023YFA1009600).
}
   }
                     
\maketitle

\begin{abstract}
Coset graphs provide a natural connection between linear codes and graph theory. In this paper, we investigate the results from the coset graph, and use spectral, structural, and probabilistic properties of coset graphs to derive relationships and bounds for the structure of linear codes and their duals. We develop bounds based on closed walks, non-backtracking walks, exponential kernels, Cheeger inequalities and random walks. We further extend the framework to $q$-ary codes and recover the $q$-ary first MRRW bound. Finally, we introduce a quotient Gilbert graph, leading to refinements of the Gilbert–Varshamov bound. Overall, the coset-graph viewpoint provides a unified graph-theoretic framework for studying the the parameters of codes.
\end{abstract}

\section{Introduction}

Error-correcting codes provide the mathematical foundation for reliable communication over noisy channels. By introducing structured redundancy, they enable the detection and correction of errors and have become essential in communication systems, storage devices, distributed computing, and information theory.

The maximum number of codewords with given length and minimum Hamming distance is a hard and classical problem in coding theory. Therefore, a central objective of coding theory is to establish finite-length and asymptotic bounds. The Gilbert-Varshamov (GV) bound provides a general lower bound \cite{Gilbert1952comparison, Varshamov1957Estimate} and remains the best known asymptotic lower bound to date. Classical upper bounds include the Hamming, or sphere-packing, bound \cite{Hamming1950Error}; the Singleton bound \cite{Singleton1964Maximum}; and the Plotkin bound \cite{Plotkin1960Binary}. These bounds are effective in complementary parameter regimes and serve as standard benchmarks for code performance \cite{MacWilliams1977theory}.

Delsarte formulated this problem within the framework of association schemes, describing the code structure by its distance distribution and using the Krawtchouk polynomials to obtain linear constraints \cite{Delsarte1973algebraic}. Optimizing over these constraints yields the Delsarte linear-programming bound. McEliece, Rodemich, Rumsey, and Welch subsequently derived the classical MRRW asymptotic bounds \cite{McEliece1977New}. This bound has been optimal for nearly fifty years. Recently, OpenAI published new binary and spherical code bounds, which strictly improve the MRRW bound in \cite[Chapter 2]{openai2026tenadvances}. The bound was soon improved again by \cite{gay2026honeycomb}, with the help of GPT-5.6 Pro.

Graph-theoretic methods play an critical roles in the improvement of GV bound. Jiang and Vardy used local sparsity to improve binary code-size lower bounds \cite{JiangVardy2004}, and was extended by  Vu and Wu to $q$-ary codes \cite{VuWu2005}. Ye et al. developed a spectral approach and a linear-code construction algorithm \cite{Ye2023} to improve GV bound. 

Coset graphs provide a natural bridge between coding theory and graph theory. It associates a graph with a code and translate coding parameters into graph-theoretic quantities.  The coset graph associated has the cosets of the linear code as its vertices. Two vertices are adjacent when representatives of the corresponding cosets differ in one coordinate. An early application of coset graphs in coding theory was given by Delorme and Solé \cite{delorme1991diameter}. They used spectral properties of the graph to study the covering radius of the code. The systematic applications of this viewpoint was developed by Friedman and Tillich \cite{Friedman2005Generalized}. They extended ideas related to the Alon--Boppana theorem related to the second largest 
eigenvalue of coset graphs. The resulting spectral estimate recovers the first MRRW bound for binary linear codes. 

Based on the spectral approach of Friedman and Tillich, Iceland and Samorodnitsky \cite{Iceland2015Coset} produced rate-versus-distance upper bounds that are strictly stronger than the first MRRW bound for low-density parity-check (LDPC) codes. They used the observation that a ball in the coset graph for LDPC codes is substantially smaller than a Hamming ball of the same radius. The results were improved in \cite{shangguan2026improved} by a better estimation of the ball size. The same authors \cite{iceland2020structured} introduced discrete Ricci curvature into the study of coset graphs of locally correctable and locally testable codes. This approach yields bounds on the covering radius of these structured codes. 

Coset graphs also play an important role in the theory of completely regular codes. Completely regular codes are closely related to distance-regular graphs through the construction of coset graphs \cite{Brouwer1989,borges2014new,Borges2019,Shi2022}.

\subsection{Main Results}

The central object of this paper is the coset graph associated with a linear code and its dual code. Its vertices represent cosets of the dual code, while its adjacency spectrum is determined by the weight distribution of the original code. This correspondence provides a common starting point for investigating coding-theoretic constraints through graph theoretic methods.

Our investigation proceeds along two main directions. The first derives bounds on coding parameters from spectral and geometric properties of coset graphs. The second uses quotient Gilbert graphs to obtain lower bounds through the identification of overlapping Hamming balls. The main results are organized according to the mathematical mechanisms underlying these two directions. Throughout this paper, all results hold analogously for dual codes.

1. Trace and moment methods

By comparing closed walks in a coset graph with those in the Hamming hypercube, we obtain constraints for the code weight spectrum. These identities lead to relations among the rate, minimum distance, and dual distance and give a linear-programming formulation for upper-bounding the size of binary linear codes. We also formulate an analogous trace inequality using the non-backtracking matrix, replacing ordinary closed walks by non-backtracking walks.

Replacing powers of the adjacency matrix by the exponential kernel, we obtain an exponential-moment inequality for the weight distribution, which provides a continuous analogue of the closed-walk trace method.

2. Cheeger inequalities and random walks.

Cheeger inequalities allow several disjoint vertex subsets with small edge boundaries to be considered simultaneously. This makes it possible to obtain constraints involving the number of low-weight codewords, rather than only the minimum distance.

Applying the Cheeger inequality to vertex subsets of a coset graph converts small edge boundaries into upper bounds on the minimum distance. When a Hamming ball embeds injectively into the quotient graph, this gives explicit bounds involving rate, minimum distance, and dual distance. Higher-order Cheeger inequalities further control the ordered codeword weights and hence provide information not only about the minimum distance but also about the multiplicity of low-weight codewords.

We also introduce a random walk on the coset graph. This yields inequalities relating the complete weight distribution of linear codes to the total variation (TV) distance from the stationary distribution and to the endpoint distribution of the corresponding random walks. This yields the rate-distance constraint.

3. Extension to closed-walk framework to $q$-ary codes.

We extend the closed-walk linear-programming framework to $q>2$. Following an approach similar to that in \cite{Friedman2005Generalized}, we obtain asymptotic estimates for the number of closed walks of length $\ell$ in the $q$-ary Hamming hypercube. Combining these estimates with the generalized Alon–Boppana method recovers the $q$-ary first MRRW bound within the coset-graph framework.

4. A coset-graph refinement of the Gilbert–Varshamov bound.

We consider a Gilbert graph whose vertices are cosets of a linear code. Its maximum degree depends on the number of cosets intersecting a Hamming ball. Consequently, the maximum code size can be lower-bounded by its independence number. The classical GV bound appearing as the special case. Overlap among Hamming balls can reduce the degree and therefore improve the resulting lower bound. Several explicit constructions are considered, and an iterative procedure is proposed for enlarging the auxiliary linear code while reducing the degree of the corresponding Gilbert graph.

\subsection{Outline}
The rest of this paper is organized as follows. Section II introduces the necessary background on linear codes, classical coding bounds, coset graphs, and necessary graphical tools. In Section III, we develop the trace and moment methods for coset graphs. The Cheeger inequalities and syndrome random walks are discussed in Section IV. Section V extends the closed-walk framework to $q$-ary codes and derives asymptotic rate-versus-distance bounds. In Section VI, we introduce a quotient Gilbert graph and develops refinements of the GV bound. Section VII concludes the paper.

\section{Preliminaries}

\subsection{Basic Definitions}

In this subsection, we introduce some definitions and notations in coding theory. For more details, please refer to \cite{MacWilliams1977theory}.

Let $q$ be a prime power and $\F_q$ be the finite field with $q$ elements. A $q$-ary code $C$ with code length $n$ is a subset of the $n$-dimensional vector space $\Fqn$. The rate of $C$ is $R(C)=\frac{1}{n} \log_q|C|$. If $C$ is a linear space of $\Fqn$, $C$ is called a linear code. The dimension of $C$ is $k(C) =  \log_q|C|$.

The Hamming weight of a vector $c\in \Fqn$ is $\wt(c) = |\{i|c_i \neq 0\}|$. The Hamming distance between two vectors $c, c'\in \Fqn$ is $d(c,c') = |\{i|c_i \neq c_i'\}|$. For two sets $S,S'\subseteq \Fqn$, $d(S,S') = \min_{c\in S, c'\in S'} d(c,c')$. 

The distance distribution of $C$ is $\left(A_{i}(C)\right)_{i=0}^{n}$, where
$$
A_{i}(C)=\frac{1}{|C|}|\{((c,c') \in C^2 \mid d(c,c')=i\}|.
$$
The minimum distance of $C$, denoted as $d(C)$, is the smallest $i\neq 0$ such that $A_i\neq 0$. Denote $A_q(n,d)$ to be the maximum size of code with length $n$ and minimum distance $d$, and $A^{\text{lin}}_q(n,d)$ to be the maximum size of linear code with length $n$ and minimum distance $d$. 

The relative distance of $C$ is then given by $\delta(C) = d(C)/n$. We say a code family $(C_n)$ has rate $R$ and relative distance $\delta$ if
\[
 \frac{\log_q|C_n|}{n}\rightarrow R,
 \qquad
 \frac{d(C_n)}{n}\rightarrow\delta.
\] 

Let 
$$
R_{\max,q}(\delta)=\varlimsup_{n \rightarrow \infty} \max \left\{R(C) \mid \delta(C) \geq \delta, C \subseteq \Fqn\right\}
$$
be the maximum rate for a code family with relative distance $\delta$. 

The covering radius of $C$ is
$$
\rho(C)=\max_{x\in\mathbb{F}_{q}^{n}} \min_{c\in C} \{d(x, c)\}.
$$

The dual code $C^{\perp}$ of linear code $C$ is
$$
C^{\perp}=\left\{x \in\mathbb{F}_{q}^{n} \mid \langle x, c \rangle =0 \quad \forall c \in C\right\}.
$$
where $\langle \cdot, \cdot \rangle$ represents inner product, defined as $\langle x, y \rangle = \sum_{i=1}^n x_iy_i$. 

The external distance of $C$ is
$$
s(C)=\bigl|\{j>0:A_j(C^{\perp})\neq 0\}\bigr|,
$$
that is, the number of distinct nonzero weights appearing in $C^{\perp}$.

\begin{definition}
For positive integers $q,t,n$, Krawtchouk polynomials are defined as
\begin{equation}
K_t(x) = \sum_{j=0}^k (-1)^j \binom{x}{j} \binom{n-x}{t-j}(q-1)^{t-j},
\end{equation}
where 
\[\binom{x}{j}=\frac{x(x-1)...(x-j+1)}{j!}.\]
Specially, $\binom{x}{0} = 1$ and $\binom{x}{j} = 0$ for $x<j$.
\end{definition}

\begin{theorem}[MacWilliams Identities]\cite{MacWilliams1977theory}
$$
A_t(C^{\perp}) =  \frac{1}{|C|} \sum_{i=0}^{n} A_{i}(C) K_t(i).
$$
\end{theorem}

Define $B_q(x,t) = \{y\in \Fqn \mid d(x,y)\leq t\}$ to be the Hamming ball with radius $t$ and 
\[V_q(n,t)=| B_q(x,t)| = \sum_{i=0}^t \binom{n}{i}(q-1)^i.\]
to be be the number of vectors with Hamming weight less than $t$.

\begin{theorem}[Gilbert-Varshamov Bound\cite{Gilbert1952comparison,Varshamov1957Estimate}]\label{pro:1.1}
\begin{equation}
A_q(n, d)\geq \frac{q^n}{V_q(n,d-1)}.
\end{equation}
\end{theorem}

For $0\le x\le1-1/q$, the $q$-ary entropy function is denoted as 
\[
 H_q(x)=x\log_q(q-1)-x\log_qx-(1-x)\log_q(1-x).
\]
Then $V_q(n,t) = q^{H_q(t/n)n + o(n)}$. The Gilbert–Varshamov bound provides the best known asymptotic lower bound on the maximum achievable code rate.
\begin{theorem}[Asymptotic Gilbert-Varshamov Bound\cite{Gilbert1952comparison,Varshamov1957Estimate}]\label{pro:1.2}
For every $0\leq \delta<1-1/q$, 
\begin{equation}
R_{\max,q}(\delta)\geq 1-H_q(\delta).
\end{equation}
\end{theorem}

\begin{theorem}[Delsarte's linear programming bound\cite{Delsarte1973algebraic}]
$$
\begin{aligned}
A_q(n, d) & \leq \max \{ \sum_{i=0}^{n} A_{i} \mid A_{0}=1,  A_{i}=0 \text { for } 1 \leq i<d \\
& A_{i} \geq 0, \sum_{i=0}^{n} A_{i} K_{k}(i) \geq 0 \text { for } 0\leq k \leq n\} .
\end{aligned}
$$

The dual linear programming is
$$
\begin{aligned}
A_q(n, d) &  \leq \min\{ \frac{1}{\lambda_0} \sum_{j=0}^n \lambda_j K_j(0)\mid \lambda_0>0, \lambda_j \geq 0 \\
&  \text { for } 1 \leq j\leq n,
\sum_{j=0}^n \lambda_j K_j(i) \leq 0 \text { for } d\leq i \leq n\}.
\end{aligned}
$$

Every feasible dual choice gives an upper bound on $A_q(n,d)$.
\end{theorem}

The celebrated upper bound is from McEliece, Rodemich, Rumsey and Welch:

\begin{theorem}[MRRW-I Bound \cite{McEliece1977New}]
\label{ThmMRRW}
$$
R_{\max,q}(\delta) \leq H_q\left(\frac{q-1-(q-2) \delta-2 \sqrt{(q-1) \delta(1-\delta)}}{q}\right)
$$
where $0<\delta<1-\frac{1}{q}$.
\end{theorem}

\begin{theorem}[MRRW-II Bound \cite{McEliece1977New}]
$$
R_{\max,2}(\delta) \leq \min_{0 \leq u \leq 1-2 \delta}\left\{1+g\left(u^2\right)-g\left(u^2+2 \delta u+2 \delta\right)\right\} .
$$
where $0<\delta<1 / 2$, $g(x)=H_2\left(\frac{1-\sqrt{1-x}}{2}\right)$.
\end{theorem}

In the following part, when $C$ and $C^{\perp}$ are clear, we abbreviate $d(C)$ as $d$ and $d(C^{\perp})$ as $d^{\perp}$. Other notations are similar.

\subsection{Coset Graph}

In this section, we introduce the definition, properties and previous works for coset graph. For more information on the basic knowledge of graph theory, please refer to \cite{Gross2013handbook}.

\begin{definition}
The coset graph $\GCp$ is the graph with vertex set $V(\GCp) = \{x+C^{\perp}|x\in \Fqn\}$ and edge set $E(\GCp) =\{(x+C^{\perp}, y+C^{\perp}) \mid x,y\in \Fqn, d(x+C^{\perp},y+C^{\perp}) = 1\}$.  
\end{definition}

When $C^{\perp}$ only contains all-zero vector, $\GCp$ is the q-ary Hamming graph $\mathcal{H}_q$. In general, $\GCp$ is quotient graph $\mathcal{H}_q / V(\GCp)$. 

When $d(C^\perp)>2$, $\GCp$ is an $n(q-1)$-regular simple graph with $q^k$ vertices.  It is a Cayley graph on the quotient group $\Fqn / C^{\perp}$ with difference set $\{ae_i\mid i=1,...,n, a\in \F_q^*\}$. Here, $e_i$ is the vector with 1 at the $i$-th coordinate and 0 otherwise. To simply the description, we only consider the code $C$ with $d(C^{\perp})>2$. 

In $\GCp$, every directed edge can be represented as $ae_i$, where $i\in{1,\ldots,n}$ and $a\in\F_q^*$. When only the direction of a closed walk matters, we denote it by $(a_1e_1,\ldots,a_te_t)$. We have $\sum_{i=1}^t a_ie_i \in C^{\perp}$.

In general, $\GCp$ is vertex transitive but not edge transitive, but if $C^{\perp}$ is transitive (e.x., a cyclic code), then $\GCp$ is arc transitive. 

Define $A(\GCp)$ to be the adjacency matrix of $\GCp$ and $L(\GCp)$  to be the normalized Laplacian matrix of $\GCp$, i.e., $L(\GCp) = I - \frac{A(\GCp)}{n(q-1)}$. Let $\lambda_{1}(C) \geq \lambda_{2}(C) \geq \cdots \geq \lambda_{q^k}(C)$ be the eigenvalues of $A(\GCp)$ and $\mu_{i}(C) = 1 - \frac{\lambda_{i}(C)}{n(q-1)}$ be the eigenvalues of $L(\GCp)$.

\begin{theorem}\cite[pp. 135-136]{Calderbank1985} \label{thm:eigen}
The eigenvalues of $A(\GCp)$ are $\{n(q-1)-q\wt(c)\mid c\in C\}$, and the eigenvalues of $L(\GCp)$ are $\{\frac{q\wt(c)}{n(q-1)} \mid c\in C\}$. The corresponding eigenvectors are $v_c$ with $\chi_c(x+C^{\perp}) = (\chi{\langle c, x\rangle})$ where $\chi$ is a nontrivial additive character of $\F_q$.

In other words, The eigenvalues of $A(\GCp)$ are $n(q-1)-qi$ with multiply $A_i(C)$.
\end{theorem}

In particular, $\lambda_{1}(C) = n(q-1)$ and $\lambda_{2}(C)=n(q-1)-q d(C)$, so the minimum weight of nonzero codewords in $C$ determines the spectral gap of the graph. For $q=2$, $\GC$ is bipartite if and only if $\lambda_{2^k}(C^{\perp}) = -n$, which is equal to $C$ contains all-one codeword $\mathbf{1}$.

The diameter of a graph $\mathcal{G}$ is the smallest integer $D$ such that any pair of vertices is connected by a path of length at most $D$. The diameter of $\GCp$ is $\rho(C^{\perp})$.

For $q=2$, $\mathbf{1}\notin C$ and $g$ be the length of shortest odd cycle in $\GCp$, then $g = d(C^{\perp})$ if $d(C^{\perp})$ is odd, and $g \geq d(C^{\perp})$ if $d(C^{\perp})$ is even. Indeed, for odd $t$, $(e_{i_1},...,e_{i_t})$ is a cycle in $\GCp$ if and only if $\sum_{j=1}^t e_{i_j}$ is a non-zero codeword in $C^{\perp}$. 

By Theorem \ref{thm:eigen}, $s(C) + 1$ is the number of distinct eigenvalues of $\GC$, together with the fact that if $\mathcal{G}$ is a graph with diameter $D$, then $\mathcal{G}$ has at least $D+1$ distinct eigenvalues,  we can imply Delsarte's external-distance principle:
$$
\rho(C)\leq s(C).
$$

The authors in \cite{delorme1991diameter} derived an upper bound on the covering radius of a code as a function of the weights of the dual code base on the idea of coset graph. We shall say that $\theta$ is a width for the code $C$ if $
\forall x \in C^{\perp}, |n(1-1 / q)-w(x)| \leqslant \theta.$

\begin{theorem}\cite[Theorem 9.]{delorme1991diameter}
Let $C$ be a code of length $n$ and width $\theta$ and dimension $K$ over $\mathbb{F}_q$ such that either $q>2$ or $\mathbf{1} \notin C^{\perp}$. Then its covering radius $\rho(C)$ satisfies

$$
\rho(C) \leqslant \left\lceil\frac{n-K}{\log _q(n(q-1) / q \theta)}\right\rceil
$$
\end{theorem}

The authors in \cite{Friedman2005Generalized} extended Alon Bopanna Theorem to prove binary MRRW-I bound (Theorem \ref{ThmMRRW}).

\subsection{Graph Tools}\label{sec:GT}

From Stirling's formula,
\[
 (n+1)^{-1}e^{nh_2(k/n)}\le\binom{n}{k}\le e^{nh_2(k/n)},
\]

A walk of length $k$ in a graph is a sequence of vertices $v_0, v_1, v_2, \ldots, v_k$ such that each consecutive pair $v_{r-1}$ and $v_r$ is connected by an edge. A closed walk of length $k$ is a walk whose starting vertex and ending vertex are the same: $v_0=v_k$.
 
Let $A$ be an $n \times n$ square matrix. The trace of $A$, denoted by $\operatorname{tr}(A)$, is the sum of all eigenvalues of $A$. For a graph with adjacency matrix $A$, the trace of the $k$-th power of the adjacency matrix $\operatorname{tr}\left(A^k\right)$ is the sum of the total number of closed walks of length $k$ in the graph.

Let $G=(V,E)$ be a finite, undirected graph with $|V|$ vertices and $|E|$ edges. Each undirected edge $\{u,v\}\in E$ is replaced by two directed edges $(u\to v)$ and $(v\to u)$. 

The non-backtracking matrix, also known as the Hashimoto
matrix\cite{hashimoto1989zeta}, is the $2|E|\times 2|E|$ matrix $B$ indexed by directed edges and defined by
\[
B_{(u\to v),(u'\to v')}
=
\begin{cases}
1, & \text{if } v=u' \text{ and } v'\neq u,\\
0, & \text{otherwise}.
\end{cases}
\]
Thus, $B_{e,f}=1$ if and only if the directed edge $f$ can follow $e$ without immediately traversing the same edge in the reverse direction. Consequently, powers of $B$ count non-backtracking walks on the graph.

\begin{theorem}\cite{bass1992ihara}
For regular graph $G=(V,E)$ with degree $D$, suppose that $\lambda$ is an eigenvalue of the adjacency matrix $A$, then each $\lambda$ corresponds to two eigenvalues $\gamma$ in non-backtracking matrix $B$ satisfy
\[
\gamma^2-\lambda\gamma+(D-1)=0.
\]
Therefore,
\[
\gamma
=
\frac{\lambda\pm\sqrt{\lambda^2-4(D-1)}}{2}.
\]

The remaining nontrivial eigenvalues of $B$ are $\pm 1$ with multiplies $|E|-|V|$.
\end{theorem}

For a graph $G=(V,E)$, $S\subseteq V$, denote $\partial S \subseteq E$ to be the set of edges from $S$ to $V \setminus S$. 

\begin{theorem}[Cheeger Inequality]\label{prop:cut-bound}
For a $D$-regular graph $G=(V,E)$, $S\subseteq V$,
\[
        \mu_2(G)
        \leq
        \frac{|V||\partial S|}{D|S|(|V|-|S|)}.
\]
\end{theorem}

\begin{theorem}[High-order Cheeger Inequality]\label{prop:ho-cut-bound0}
For a $D$-regular graph $G=(V,E)$, $m\geq 2$,
\[
        \frac{\mu_m(G)}{2}
        \leq
        \min_{S_1,\dots,S_m}
        \max_{1\leq j\leq m}
        \frac{|\partial S_j|}{D|S_j|},
\]
where the minimum is over $m$ nonempty pairwise disjoint subsets of $V = \{S_1,...,S_m\}$. 
\end{theorem}

Consider an operator $f$ on finite Abelian group $G$. Define $\chi_x$ to be characters of $G$. The Fourier transformation of $f$ is
$$
\hat{f}(y)=\sum_{x \in G} f(x) \bar{\chi_y(x)}.
$$
Then
$$
f(x) = \frac{1}{|G|} \sum_{y\in G} \hat{f}(y) \chi_x(y).
$$

\begin{theorem}[Parseval's identity]
$$
\sum_{x \in G} f(x) \overline{g(x)}= \frac{1}{|G|} \sum_{c \in \widehat{G}} \hat{f}(c) \overline{\hat{g}(c)}
$$
\end{theorem}

Consider the random walk on the group $G$. At each step, the probability from $y\to x+y$ is denoted as $p(x)$. The transition matrix is denoted as $P$. Then
\begin{align*}
& (P\chi_x)(z)=\sum_y p(y) \chi_x(z+y) \\
 = & \chi_x(z) \sum_y p(y)  \chi_x(y) = \hat{p}(x) \chi_x(z)
\end{align*}
Therefore, $\hat{p}(x)$ is the eigenvalue of $P$.

\subsection{Puncturing, shortening and direct sum}

In this subsection, we introduce the coset graph after puncturing, shortening and direct sum.

\begin{proposition}

a) If $C_1$ is obtained by puncturing $C$ at coordinate $i$, then $\mathcal{G}_{C_1^\perp}$ is isomorphic to each connected component of the graph obtained from $\GCp$ by deleting all coordinate-$i$ edges. If puncturing preserves dimension, the remaining graph is connected.

b) If $C_2$ is obtained by shortening $C$ at coordinate $i$, then $\mathcal{G}_{C_2^{\perp}}$ is obtained by contracting all coordinate-$i$ edges.

c) Let $\oplus$ denote the direct sum of two codes.
\[
 \mathcal{G}_{C\oplus C'}
 \cong\GC\mathbin{\square}\mathcal{G}_{C'},
\]
where $\square$ is the Cartesian graph product.  
\end{proposition}

\begin{proof}

If we remove all edges labelled $i$, the graph splits into components. Each component is a coset of $\operatorname{span}\{ae_j:j\ne i, a\in\F_q^*\}$. These components form the coset graph of the punctured code.

For shortening, the cosets $S$ and $ae_i+S$ merge into one. This is exactly the same as contracting all edges labelled $i$, while keeping the other coordinate labels unchanged.

For direct sum, every coset of $C\oplus C'$ is uniquely of the form $(x+C,x'+C')$. Moreover, two such cosets are adjacent exactly when their difference has Hamming weight 1, which occurs precisely when one coordinate defines an edge in the corresponding coset graph and the other coordinate is unchanged. This is exactly the adjacency rule for the Cartesian product. Hence the two graphs are isomorphic.

\end{proof}

\section{Trace and Moment Bounds for Binary Codes}

\subsection{Trace bound and linear programming bounds}

In this and next subsection, we consider the case $q=2$.

Let $N_\ell$ be the number of ordered coordinate sequences $(i_1,\ldots,i_\ell)\in[n]^\ell$ in which every coordinate occurs an even number of times. This is the number of length-$\ell$ closed walks from a fixed vertex of $\mathcal H_{n,2}$. Homomorphism gives at least $2^kN_\ell$ closed walks in $\GCp$ \cite{Friedman2005Generalized}. Extra walks project to nonzero words of $C^\perp$. 

Since the number of closed walks with length $\ell$ is equal to the trace of $\ell$-th order adjacency matrix, we have
$$
\sum_{i=0}^n A_i(C)(n-2i)^{\ell}\geq 2^k N_{\ell}.
$$
When $\ell$ is odd, $N_{\ell} = 0$, then
$$
\sum_{i=0}^n A_i(C)(n-2i)^{\ell}\geq 0.
$$
Also, when $\ell < d(C^{\perp})$, 
\begin{equation}\label{eq:lp0}
\sum_{i=0}^n A_i(C)(n-2i)^{\ell} =2^k N_{\ell}.
\end{equation}
In other words, if $\ell < d(C^{\perp})$, since $1 + \sum_{i=d}^n A_i(C) = |C|$, 
\begin{equation}\label{eq:lp}
\sum_{i=d}^n A_i(C)((n-2i)^{\ell}-N_{\ell})= N_{\ell}-n^{\ell}.
\end{equation}

Note that $N_{\ell}$ is exactly the number of closed walks per vertex in hypercube $\mathcal{H}_2$. 

\begin{theorem}\cite{Friedman2005Generalized}\label{thm:Friedman}
$$
N_{\ell}=\frac{1}{2^n} \sum_{i=0}^n\binom{n}{i}(n-2 i)^{\ell}.
$$
Moreover, $N_\ell=0$ for odd $\ell$. For even integers $\ell=\alpha n+O(1)$ with fixed $\alpha>0$,
\begin{equation}\label{Friedman_asy}
N_\ell=n^\ell 2^{n[H_2(\beta_\alpha)-1+\alpha\log_2(1-2\beta_\alpha)]+o(n)},
\end{equation}
where $\beta_\alpha\in(0,1/2)$ is the unique solution of
\[
(1-2\beta_\alpha)\ln(\beta_\alpha^{-1}-1)=2\alpha.
\]
\end{theorem}

Next, we provide two simple propositions, by simply looking at one equation in \eqref{eq:lp0} for even and odd $\ell$ receptively.

\begin{proposition}\label{por2}
For a positive even integer $\ell<d^\perp$,
\[
2^k\geq\frac{n^\ell}{N_\ell}.
\]
For a binary linear-code family with rate $R$ and dual relative distance $\delta^\perp>0$, every $0<\alpha<\delta^\perp$ gives
\[
R\geq1-H_2(\beta_\alpha)-\alpha\log_2(1-2\beta_\alpha).
\]
The same inequality holds at $\alpha=\delta^\perp$ by continuity.
\end{proposition}
\begin{proof}
Equation~\eqref{eq:lp0} and the nonnegativity of even powers imply
\[
2^kN_\ell=n^\ell+\sum_{i=1}^nA_i(C)(n-2i)^\ell\geq n^\ell.
\]
Choose even $\ell_n=\alpha n+O(1)$ with $\alpha<\delta^\perp$, take base-two logarithms, divide by $n$, and apply~\eqref{Friedman_asy}.
\end{proof}

\begin{remark}
It can be calculated that the bound in Proposition \ref{por2} is better than Plotkin and Elias–Bassalygo bound when $\delta$ is large, but always weaker than MRRW bound.
\end{remark}

\begin{proposition}\label{por3}
For a positive odd integer $\ell<d^\perp$,
\[
\max_{c\in C\setminus\{0\}}\frac{|n-2\wt(c)|}{n}
\geq(2^k-1)^{-1/\ell}.
\]
Furthermore, if all-one codeword does not belong to $C$,
$$
1 -  \frac{2}{n}\lceil \frac{d(C)}{2} \rceil \geq  (2^k-1)^{-1/\ell}.
$$
\end{proposition}
\begin{proof}
Equation~\eqref{eq:lp0} gives
\[
n^\ell+\sum_{c\in C\setminus\{0\}}(n-2\wt(c))^\ell=0.
\]
Taking absolute values yields
\[
n^\ell\leq(2^k-1)\max_{c\in C\setminus\{0\}}|n-2\wt(c)|^\ell.
\]

If all-one codeword does not belong to $C$, then the codewords in $C$ have weight at most $n - \lceil \frac{d(C)}{2} \rceil$. Thus, 
$$
\max_{c\in C\setminus\{0\}} |n-2\wt(c)| \geq n - 2\lceil \frac{d(C)}{2} \rceil.
$$
\end{proof}

Proposition~\ref{por3} is a two-sided spectral constraint, of both the largest and smallest codeword weight.

From \eqref{eq:lp}, we obtain the linear programming bound.

\begin{theorem}\label{thm:LP}
$$
\begin{aligned}
A^{\text{lin}}_2(n, d) & \leq \max \{1+\sum_{i=d}^{n} A_{i} \mid   A_{i} \geq 0, i = d,\cdots,n,\\
& \sum_{i=d}^{n} A_i((n-2i)^{k}-N_k)\geq N_{k} - n^k \\
& \text { for } k= 0, \cdots, n\} .
\end{aligned}
$$

The dual linear programming is
$$
\begin{aligned}
A^{\text{lin}}_2(n, d) & \leq \min\{1+\sum_{j=0}^n \lambda_j (n^j - N_j)\mid  \\
& \lambda_j \geq 0, \quad j=0, \cdots, n, \\
& \sum_{j=0}^n \lambda_j (N_j - (n-2i)^j) \geq 1, \quad i=d, \cdots, n\}.
\end{aligned}
$$
\end{theorem}

The general trace inequalities, without a prescribed dual distance, give the following relaxation of Theorem~\ref{thm:LP}. The equalities in~\eqref{eq:lp} are available only for moment orders below $d^\perp$.

Any solution of dual linear programming is an upper bound of $A^{\text{lin}}_2(n, d) $. There is no possible solution that only one variable is not zero. If two variables are not zero, we have the following proposition.

\begin{proposition}
Let $j$ be even with $2\leq j\leq n-1$.  Put $Q=(n-2d)^j(1+n-2d)$. If $N_j>Q$, then
\begin{equation}
 A_2^{\mathrm{lin}}(n,d)
 \leq1+\frac{n^j-N_j+n^{j+1}}{N_j-Q}.
\end{equation}
\end{proposition}

\begin{proof}
In the dual linear programming , set $\lambda_j=\lambda_{j+1}=(N_j-Q)^{-1}$ and all other variables to zero.
Because $N_{j+1}=0$, the constraints for $d\leq i\leq n$ are
\[
\frac{N_j-(n-2i)^j(1+n-2i)}{N_j-Q}\geq 1,
\]
For even $j$, the function $x^j(1+x)$ is increasing on $[0,\infty)$ and on $(-\infty,-1]$. If $d\leq i\leq n/2$, then $(n-2i)^j(1+n-2i)\leq Q$.  If $ i> n/2$, then $(n-2i)^j(1+n-2i)< 0$. Thus all constraints hold, and substitution gives the stated objective value.
\end{proof}

Recall the definition of non-backtracking matrix in Section \ref{sec:GT}. For convenience, denote $\gamma_1(\lambda) = \frac{\lambda + \sqrt{\lambda^2-4(n-1)}}{2}$ and $\gamma_2(\lambda) = \frac{\lambda - \sqrt{\lambda^2-4(n-1)}}{2}$. 

\begin{theorem}\label{thm:nonbacktracking}
For $\ell\geq1$, put $p_\ell(\lambda)=\gamma_1(\lambda)^\ell+\gamma_2(\lambda)^\ell$ and define the normalized hypercube trace
\[
M_\ell=2^{-n}\sum_{i=0}^n\binom ni p_\ell(n-2i)
 +(1+(-1)^\ell)(n/2-1).
\]
For a binary linear code with $d(C^\perp)>2$,
\begin{align*}
&\sum_{i=0}^n A_i(C)p_\ell(n-2i)\\
&\quad +(1+(-1)^\ell)(n/2-1)|C|\geq |C|M_\ell.
\end{align*}
Equality holds if $\ell<d(C^\perp)$.
\end{theorem}

The similar linear programming bound as Theorem~\ref{thm:LP} can be deduced from Theorem~\ref{thm:nonbacktracking}.

The preceding bounds are obtained from individual spectral moments, each corresponding to closed walks of a prescribed length. A natural extension is to combine these moments through a generating function. The matrix exponential provides such a formulation, since its trace is a weighted sum of closed-walk counts over all lengths. This motivates the exponential-kernel approach developed in the following subsection.

\subsection{Exponential moment bounds}

The generalized Alon--Boppana argument in the coset-graph setting \cite{Friedman2005Generalized} can be viewed as a trace method using powers $A^\ell$.  A variant use is the exponential moment function $e^{tA}$ for every $t>0$,
\[
        \frac{1}{2^k}\tr\bigl(e^{tA(G_{C^{\perp}})}\bigr)
        \geq
        \frac{1}{2^n}\tr\bigl(e^{tA(\mathcal{H}_2)}\bigr).
\]
Note that 
\begin{align*}
\tr\left(e^{t A\left(\mathcal{H}_2\right)}\right) & =\sum_{k=0}^n\binom{n}{k} e^{t(n-2 k)} =e^{n t} \sum_{k=0}^n\binom{n}{k} e^{-2 t k} \\
& =e^{n t}\left(1+e^{-2 t}\right)^n  =(2 \cosh t)^n .
\end{align*}

Using the spectral description of $G_{C^{\perp}}$, we obtain the exponential moment inequality
\[
        \frac1{2^k}
        \sum_{c\in C}
        e^{t(n-2\wt(c))}
        \geq
        (\cosh t)^n.
\]
Separating the zero codeword gives
\[
        \sum_{c\in C,\,c\neq 0}
        e^{t(n-2\wt(c))}
        \geq
        2^k(\cosh t)^n-e^{tn}.
\]
Since every nonzero codeword has weight at least $d(C)$,
\[
        \sum_{c\in C,\,c\neq 0}
        e^{t(n-2\wt(c))}
        \leq
        (2^k-1)e^{t(n-2d(C))}.
\]
Therefore, 
\[
        (2^k-1)e^{t(n-2d(C))}
        \geq
        2^k(\cosh t)^n-e^{tn}.
\]
To take the logarithm of the resulting lower bound, its right-hand side must be positive. Define 
$$
T = \{t>0:2^k(\cosh t)^n>e^{nt}\} = \left(0,\frac12\ln\frac1{2^{1-k/n}-1}\right),
$$ 
we get
\[
        d(C)\leq
        \frac{n}{2} - \frac{1}{2t}
        \log\left(
        \frac{2^k(\cosh t)^n-e^{tn}}{2^k-1}
        \right).
\]
Optimizing over $t>0$ gives the one-parameter bound

\begin{theorem}
\[
        d(C)\leq
        \inf_{t\in T}
        \left[
        \frac{n}{2} - \frac{1}{2t}
        \log\left(
        \frac{2^k(\cosh t)^n-e^{tn}}{2^k-1}
        \right)\right].
\]
\end{theorem}

\section{Expansion and Random-Walk Bounds}

This section applies standard Cheeger and Fourier inequalities to coset graphs. The finite-length consequences retain dual-distance, ordered-weight, or support information.

\subsection{Cheeger cuts and bounds}

In this section, we first consider the Cheeger inequalities for binary case, then we generalize it to $q$-ary and higher-order Cheeger inequalities.

For the coset graph $\GCp$, recall that $\mu_2 = 2d/n$, by Theorem \ref{prop:cut-bound}, we obtain the following direct coding consequence.

\begin{proposition}\label{prop:cut-bound1}
\[
        d\leq\frac{2^k|\partial S|}{2|S|(2^k-|S|)}.
\]
In particular, if $|S|\leq 2^{k-1}$, then
\[
        d\leq \frac{|\partial S|}{|S|}.
\]
\end{proposition}

This gives a simple principle: any vertex set in the coset graph with small edge boundary immediately yields an upper bound on the minimum distance of $C$.

\begin{theorem}\label{cor:cheeger-dual}
Let
\[
        r=\left\lfloor \frac{d(C^{\perp})-1}{2}\right\rfloor.
\]
If $V_2(n,r)<2^k$, then
\[
        d(C)
        \leq
        \frac{
            2^k (n-r)\binom{n}{r}
        }{
            2V_2(n,r)\bigl(2^k-V_2(n,r)\bigr)
        }.
\]
If, moreover, $V_2(n,r)\leq 2^{k-1}$, then the cleaner estimate
\[
        d(C)
        \leq
        \frac{(n-r)\binom{n}{r}}{V_2(n,r)}
\]
holds.
\end{theorem}

\begin{proof}
Consider a Hamming ball of radius $r$ in $\F_2^n$.  If       $r<d^{\perp}/2$, then the natural quotient map $\F_2^n\to \F_2^n/C^{\perp}$ is injective on the Hamming ball of radius $r$.  Hence the radius-$r$ ball around the zero coset in $\GCp$ has exactly $V_2(n,r)$ vertices. There are $\binom{n}{r}$ vertices at the boundary, and each vertex has at most $(n-r)$ edges connected to outside. Combining this with Proposition \ref{prop:cut-bound1} gives the following explicit relation among $d$, $d^{\perp}$, $k$, and $n$.
\end{proof}

\begin{remark}
Theorem~\ref{cor:cheeger-dual} does not beat classical asymptotic coding bounds, but it is obtained from a very transparent graph-theoretic object. It is therefore well suited to refinements using additional structure, for example cyclic symmetry, local constraints, or irregular quotient partitions.
\end{remark}

\subsubsection{Higher-order Cheeger inequalities and low-weight spectra}

List the weights of codewords in $C$ with multiplicity as
\[
        0=w_1\leq w_2\leq \cdots \leq w_{q^k},
\]
By Theorem \ref{thm:eigen} and \ref{prop:ho-cut-bound0},

\begin{proposition}\label{prop:higher-cheeger}
If $G=\GCp$ contains $m$ pairwise disjoint sets $S_1,\dots,S_m$ satisfying
\[
        \frac{|\partial S_j|}{|S_j|}\leq h
        \qquad\text{for all }j,
\]
For every $m\geq 2$,
\[
        \frac{qw_m(C)}{2n(q-1)}\leq \frac{h}{n(q-1)}.
\]
Consequently, 
\[
        w_m(C)\leq \frac{2h}{q}.
\]
\end{proposition}

A concrete consequence is obtained by taking disjoint Hamming balls inside the coset graph.

\begin{theorem}\label{cor:many-balls}
Assume $2r<d(C^{\perp})$ and
\[
        mV_q(n,2r)\leq q^k.
\]
Then
\[
        w_m(C)
        \leq
        \frac{2(n-r)\binom{n}{r}(q-1)^{r+1}}{qV_q(n,r)}.
\]
\end{theorem}

\begin{proof}
The condition $mV_q(n,2r)\leq q^k$ allows one to greedily choose $m$ centers in $\GCp$ whose pairwise distances are greater than $2r$.  The radius-$r$ balls around these centers are disjoint.  Since $2r<d(C^{\perp})$, each such ball has exactly $V_q(n,r)$ vertices. Every ball has vertex boundary with size $\binom{n}{r}(q-1)^r$, and each vertex has outer degree at most $(n-r)(q-1)$ . Then apply Proposition \ref{prop:higher-cheeger}.
\end{proof}

This result does not only bound the first nonzero weight $w_2=d(C)$, but also controls how many low-weight codewords must occur.

\subsection{Random Walks}

The expansion inequalities considered above provide geometric constraints on coset graphs. We now examine the same graphs from a probabilistic perspective. Since the transition operator of the natural random walk is a normalized adjacency operator, its eigenvalues are directly related to the weight distribution of the underlying code. This allows mixing properties of the walk to be translated into coding-theoretic inequalities.

Consider the following random walk on $\GCp$.  At each step, choose $i\in[n]$ and $a\in\F_q$ independently and uniformly, and add $a e_i$ to the current coset. Equivalently, one randomly selected error coordinate is resampled uniformly.  Let $P$ be the transition matrix, then
\begin{equation}\label{eq:tran}
 P=\frac{1}{q} I+\frac{1}{qn}A(\GCp)=I-\frac{(q-1)L{(\GCp)}}{q}.
\end{equation}
Let $p_t$ be the distribution at time $t$ started from the zero coset, and $p=p_1$.  We use standard total-variation and mixing-time terminology as in \cite{levin2017markov}.

\begin{proposition}\label{thm:mixexact}
For $c\in C$, the character indexed by $c$ is a $P$-eigenfunction with
eigenvalue
\begin{equation}\label{syn1}
 \lambda_c(P) =1-\frac{\wt(c)}{n}.
\end{equation}
For every integer $t\ge0$,
\begin{equation}\label{syn2}
 q^k\sum_{x\in \GCp}\bigl(p_t(x)-\frac{1}{q^k}\bigr)^2
 =\sum_{j=1}^{n}A_j(C)\left(1-\frac jn\right)^{2t}.
\end{equation}
Moreover, if $X_t,Y_t$ are independent copies of the walk, both started from zero, then
\begin{equation}\label{syn3}
 \mathbb{P}(X_t=Y_t)=\frac{1}{q^k}\left[
 1+\sum_{j=1}^{n}A_j(C)\left(1-\frac jn\right)^{2t}
 \right].
\end{equation}

\end{proposition}

\begin{proof}
Fix a nontrivial additive character $\chi$ of $\F_q$. For the character $\chi_c(x+C^\perp)=\chi(\langle c,x\rangle)$,
\[
 P\chi_c
 =\frac1n\sum_{i=1}^n\left(\frac1q
 \sum_{a\in\F_q}\chi(a c_i)\right)\chi_c.
\]
The parenthesized average is $1$ if $c_i=0$ and $0$ otherwise, proving \eqref{syn1}.
 
Since $p_t$ is obtained by convolving the one-step distribution $p$ with itself $t$ times, 
$$
\widehat{p_t}(c)= \widehat{p}(c)^t =\lambda_c(P) ^t.
$$
Parseval's identity gives
\[
\sum_x p_t(x)^2=\frac{1}{q^k}\sum_{c\in C}\lambda_c(P)^{2t},
\]
and the left side is $\mathbb{P}(X_t=Y_t)$, proving \eqref{syn3}. \eqref{syn2} is also followed.
\end{proof}

\begin{remark}
Equation~\eqref{syn3} can be applied to estimate the maximum-likelihood decoding error probability when $\frac{(q-1)2t}{q}$ repeated coordinate errors occur.
\end{remark}

The TV mixing time measures how quickly a Markov chain approaches its stationary distribution. The TV distance at time $t$ is defined by
\[
d_{\text{TV}}(t) 
= \frac{1}{2}\sum_{x\in\GCp} |p_t(x) - 1/q^k|.
\]

For a given tolerance $\varepsilon>0$, the mixing time is
\[
t_{\text{mix}}(\varepsilon)  
= \min\left\{t\in\mathbb{Z}^+:d_{\mathrm{TV}}(t)\le \varepsilon \right\}.
\]

Thus, the total-variation mixing time is the minimum number of steps needed for the distribution of the Markov chain to become sufficiently close to its stationary distribution, even from the worst initial state\cite{levin2017markov}.

\begin{proposition}\label{pro:mixing}

\begin{equation}\label{syn4}
 \frac{q-1}q\left(1-\frac dn\right)^t
 \le d_{\text{TV}}(t) 
 \le \frac12\sqrt{q^k-1}
       \left(1-\frac dn\right)^t.
\end{equation}

Consequently, for $0<\varepsilon<1/2$, the total-variation mixing time $t_{\mathrm{mix}}(\varepsilon)$ satisfies

\begin{align}\label{syn5}
\frac{\log((q-1)/(q\varepsilon))}
      {-\log(1-d/n)}
 \le &  t_{\mathrm{mix}}(\varepsilon) \notag \\ 
 \le & \left\lceil
 \frac{\tfrac12\log(q^k-1)+\log(1/(2\varepsilon))}
      {-\log(1-d/n)}
 \right\rceil.
\end{align}

\end{proposition}

\begin{proof}
Cauchy--Schwarz inequality and \eqref{syn2} give
\[
 d_{\text{TV}}(t) 
 \le\frac{1}{2}\left(
  \sum_{j=1}^nA_j(C)(1-\frac{j}{n})^{2t}
 \right)^{1/2}.
\]
All nonzero codewords have weight at least $d$, proving the upper bound in \eqref{syn4}.  

For the lower bound, choose a codeword $c$ with weight $d$ and project the walk by the surjective map $x\mapsto\langle c,x\rangle\in\F_q$. The coordinate in the support of $c$ contributes a uniform increment $\pi$ on $\F_q$; other coordinates contribute zero. Now the probability for not selecting any coordinates in $\supp(c)$ in time $t$ is $(1-\frac{d}{n})^t$. Otherwise, once a random variable is added to the projection, the distribution will be a random uniform variable, with the probability $1-(1-\frac{d}{n})^t$.

Therefore, its TV distance to a uniformly random distribution is $\frac{q-1}{q}(1-\frac{d}{n})^t$. TV cannot increase under a map, proving the lower bound.   

Solving the two inequalities for $t$ proves \eqref{syn5}.  Translation invariance makes the bound independent of the initial syndrome.
\end{proof}

Define $B_{\Cp}(x,r)$ to be the set of cosets of $\Cp$ with distance at most $r$ from the coset $x+\Cp$. The walk is supported on this set. Combining support size with Parseval yields the following standard support--$L^2$ consequence.

\begin{theorem}\label{thm:specvol}
For every integer $t\ge0$,
\[
1-\frac{V_q(n,t)}{q^k}\leq
 1-\frac{|B_{\Cp}(0,t)|}{q^k}\leq
 \frac{1}{2}\left[
 \sum_{j=1}^{n}A_j(C)\left(1-\frac jn\right)^{2t}
 \right]^{1/2}.
\]
\end{theorem}

\begin{proof}
Since $p_t$ is supported on $B_C(0,t)$,
\begin{align*}
2d_{\text{TV}}(t) & = \sum_{x\in B_C(0,t)} |p_t(x) - 1/q^k| + \sum_{x\notin B_{\Cp}(0,t)} |p_t(x) - 1/q^k| \\
& \geq \sum_{x\in B_C(0,t)} |p_t(x)| - |B_{\Cp}(0,t)|/q^k + \sum_{x\notin B_{\Cp}(0,t)} |1/q^k| \\
& = 2(1-\frac{|B_{\Cp}(0,t)|}{q^k}).
\end{align*}
Also $|B_{\Cp}(0,t)|\le V_q(n,t)$.  The upper bound for total variation used in the proof of is the right side of Proposition \ref{pro:mixing}. Combining the two estimates proves the theorem.
\end{proof}

\begin{theorem}\label{cor:finitevolume}
Let $0<\varepsilon<1/2$, and set
\begin{equation}\label{syn6}
 t_\varepsilon=
 \left\lceil
 \frac{\tfrac12\log(q^k-1)+\log(1/(2\varepsilon))}
      {-\log(1-d/n)}
 \right\rceil.
\end{equation}
Then
\[
V_q(n,t_\varepsilon)\ge(1-\varepsilon)q^k.
\]
\end{theorem}

\begin{proof}
Proposition \ref{pro:mixing} gives $d_{\text{TV}}(t_\varepsilon)\le\varepsilon$.  Since $p_t$ is supported on
the radius-$t$ quotient ball, that ball must contain at least
$(1-\varepsilon)q^k$ vertices.  It is the image of a Hamming ball and hence has at most $V_q(n,t_\varepsilon)$ vertices.
\end{proof}

\begin{corollary}
\label{cor:asympmix}
Let $(C_n)$ be a family of $q$-ary linear codes with rate $R$ and relative distance $\delta$. Put
\[
 \tau=\frac{R\log q}{-2\log(1-\delta)}.
\]
If $\tau<1-1/q$, then
\[
R\le H_q(\tau).
\]
\end{corollary}

\begin{proof}
Fix $\varepsilon\in(0,1/2)$. Equation~\eqref{syn6} gives $t_\varepsilon/n\to\tau$. Taking base-$q$ logarithms in Theorem~\ref{cor:finitevolume} gives the result.
\end{proof}

\section{The $q$-ary Extension for closed-walk bounds}

In this section, we extend two $q$-ary extensions of the binary analysis in \cite{Friedman2005Generalized}. The first is a  closed-walk number estimate, with an additional positive/negative branch comparison for $q>2$. The second is the Dirichlet eigenvalue argument recovering the established first MRRW bound. 

\subsection{The number of closed walks}

For an integer $l\geq0$, the number of length-$l$ closed walks from vertex in $\mathcal{H}_q$ is

\begin{equation}\label{eq:Nq}
 N_{q,l}=q^{-n}\sum_{i=0}^n\binom{n}{i}(q-1)^i(n(q-1)-qi)^l
\end{equation}

Next, we are going to prove an analogue of \cite[Lemma A.2, (11)]{Friedman2005Generalized} for $q$-ary case.

Put $Q=1-1/q$. For $0\leq\beta<Q$, define
$$
 g_q(\beta)=-\beta\log\beta-(1-\beta)\log(1-\beta)
                          +\beta\log(q-1),
$$
$$                          
 F_{q,\alpha}(\beta)=g_q(\beta)-\log q +
                 \alpha\log(q-1-q\beta).
$$

For $\beta n$ and $\alpha n$ integral and $0<\beta<Q$, Stirling's formula gives
\begin{align*}
&q^{-n}\binom n{\beta n}(q-1)^{\beta n}
 \bigl(n(q-1)-q\beta n\bigr)^{\alpha n}\\
&\qquad=n^{\alpha n}\exp\{nF_{q,\alpha}(\beta)+O(\log n)\}.
\end{align*}

Differentiating the exponent gives
\[
 F'_{q,\alpha}(\beta)=\log\frac{(q-1)(1-\beta)}\beta
               -\frac{q\alpha}{q-1-q\beta},
\]
\[
 F''_{q,\alpha}(\beta)=-\frac1{\beta(1-\beta)}
               -\frac{q^2\alpha}{(q-1-q\beta)^2}<0.
\]
The first derivative tends to $+\infty$ and $-\infty$ at the two ends. Therefore, for every $\alpha>0$, $F_{q,\alpha}$ has a unique maximizer $\beta_\alpha\in(0,Q)$, determined by
\begin{equation}\label{eq:betasaddle}
 (q-1-q\beta_\alpha)
 \log\frac{(q-1)(1-\beta_\alpha)}{\beta_\alpha}=q\alpha.
\end{equation}

\begin{lemma}\label{lem:dominance}
Fix $q>2$ and $\alpha>0$. The maximum exponential rate of the
negative-eigenvalue terms in \eqref{eq:Nq} is strictly smaller than that of the positive-eigenvalue terms.
\end{lemma}

\begin{proof}
For $0<t<1/q$, put $\Delta(t)=g_q(Q-t)-g_q(Q+t)$. Then $\Delta(0)=0$ and
\[
 \Delta'(t)=\log\frac{Q^2-t^2}{Q^2-(q-1)^2t^2}>0,
\]
so $g_q(Q-t)>g_q(Q+t)$. Since $(q-1) - q(Q-t) = - ((q-1) - q(Q+t))$, every negative-branch point has a positive-branch reflected point with strictly larger exponent.
\end{proof}

\begin{theorem}\label{thm:moment}
Fix $q>2$. Let $\ell_n$ be positive integers with $\ell_n=\alpha n+O(1), \alpha\in(0,1)$. Then
\begin{equation}\label{eq:momentlog}
\log N_{q,\ell_n}=\alpha n\log n+nF_{q,\alpha}(\beta_{\alpha})+O(\log n).
\end{equation}
\end{theorem}

\begin{proof}
Write $N_{q,\ell_n}/n^{\ell_n}=P_n+(-1)^{\ell_n}T_n$, where $P_n$ is the sum of positive terms over $j<Qn$ and $T_n$ is the sum of negative terms over $j>Qn$. From Lemma~\ref{lem:dominance}, $T_n = P_n e^{O(n)}$, so it can be neglected after taking the logarithm.

For a lower bound, choose an integer $j$ maximizing $F_{q,\ell_n}(j/n)$. The continuity gives $j = n\beta_{\alpha} + O(1)$. Then
\[
 F_{q,\ell_n/n}(j/n)=F_{q,\alpha}(\beta_{\alpha})+O(n^{-1}).
\]
Therefore,
\[
 \log P_n=nF_{q,\alpha}(\beta_\alpha)+O(\log n).
\]
This proves \eqref{eq:momentlog}. 
\end{proof}

\begin{remark}
At $q=2$ the two branches are exact reflections:
$T_n=P_n$, so even moments double and odd moments cancel.
For $q=2$, Theorem \ref{thm:moment} holds along even $\ell_n$; odd moments vanish.
\end{remark}

For convenience, denote $\theta_i = n(q-1)-qi$. The linear programming is
$$
\begin{aligned}
A^{\text{lin}}_q(n, d) & \leq \max  \{1+\sum_{i=d}^{n} A_{i} \mid   A_{i} \geq 0, i = d,\cdots,n,\\
& \sum_{i=d}^{n} A_i(\theta_i^l - N_{q,l})\geq N_{q,l} - \theta_0^l \text { for } l = 0, \cdots, n\} .
\end{aligned}
$$

The dual linear programming is
$$
\begin{aligned}
A^{\text{lin}}_q(n, d) & \leq \min\{1+\sum_{j=0}^n \lambda_j (\theta_0^j - N_{q,j})\mid  \\
& \lambda_j \geq 0, \quad j=0, \cdots, n, \\
& \sum_{j=0}^n \lambda_j (N_{q,j} - \theta_i^j ) \geq 1, \quad i=d, \cdots, n\}.
\end{aligned}
$$

\subsection{The size of balls in the coset graph}

For a finite graph $G$, define
\[
 \lambda_{1,\mathrm{Dir}}(W)=
       \max_{0\ne f:\supp f\subseteq W}
                     \frac{\langle A(G)f, f\rangle}{\langle f, f \rangle},
\]
\[
 \FK_G(a)=\max_{|W|\le a}\lambda_{1,\mathrm{Dir}}(W).
\]

Next we prove the $q$-ary counterpart of \cite[Proposition 8.5]{Friedman2005Generalized}. Recall $B_q(0,r)=\{x\in\F_q^n:\wt(x)\le r\}$ and $|B_q(0,r)| = q^{H_q(r/n)n + o(n)}$.

\begin{lemma}\label{lem:ball}
For fixed $0<\tau<Q$, $r = \lfloor\tau n\rfloor$,
\begin{align}\label{eq:ball1}
\notag
& \lambda_{1,\mathrm{Dir}}(B_q(0,r)) \\ 
\geq  & n((q-2)\tau+2\sqrt{(q-1)\tau(1-\tau)})+o(n).
\end{align}
Moreover,
\begin{align}\label{eq:ball2}
\notag
& \FK_{\mathcal{H}_q}\bigl(q^{H_q(\tau) n} / \log n\bigr) \\
\geq  & n((q-2)\tau+2\sqrt{(q-1)\tau(1-\tau)})+o(n).
\end{align}
\end{lemma}

\begin{proof}
Define $v_i = \sqrt{\binom{n}{i}(q-1)^i}$. Consider a function $f$ on $B_q(0,r)$ that depend only on the weight of the vertex, and $f(x) = f_i$ for any $x$ with $\wt(x)=i$. Take $r' = r - o(1)$. If $x\notin [r', r]$, $f_i = 0$. Let $f_{\text {nrm}}$, the normalization of $f$, be the function on $[r',r]$ such that
$$
f_{\mathrm{nrm}}(i)= f_i / v_i.
$$

A weight-$i$ vertex has $i$ neighbours in layer $i-1$, $(q-2)i$ neighbours in layer $i$, and $(q-1)(n-i)$ neighbours in layer $i+1$. Therefore, for any $x$ with $\wt(x)=i$,
$$
(A(\GCp)f)(x) = i f_{i-1} + (q-2)i f_i + (q-1)(n-i) f_{i+1}.
$$

Then
\begin{align*}
& (A(\GCp)f)(x) / v_i = \sqrt{(q-1)i(n-i+1)} f_{\mathrm{nrm}}(i-1) \\ 
+ & (q-2)i f_{\mathrm{nrm}}(i) + \sqrt{(q-1)(i+1)(n-i)} f_{\mathrm{nrm}}(i+1).
\end{align*}

for all $r'<i<r$ (the coefficient of the right-hand-side vanishes for $f_{\mathrm{nrm}}$ at the values $-1$ and $n+1$). Define $\tilde{A}$ to be a tridiagonal matrix with $\tilde{A}_{i,i} = (q-2)i$ and  $\tilde{A}_{i,i+1} = \sqrt{(q-1)(i+1)(n-i)}$. We have
$$
\frac{\langle A(\GCp)f, f\rangle}{\langle f, f \rangle} = \frac{\langle \tilde{A}f_{\mathrm{nrm}}, f_{\mathrm{nrm}}\rangle}{\langle f_{\mathrm{nrm}}, f_{\mathrm{nrm}} \rangle}.
$$

Hence, 
\begin{align*}
& \lambda_{1,\mathrm{Dir}}(B_q(0,r)) \geq \lambda_1(\tilde{A})\\
\geq &  (q-2)r+2 \sqrt{(q-1)(r'+1)(n-r')} \cos \frac{\pi}{r-r'+1}  \\ 
= & (q-2)r+2 \sqrt{(q-1)(r+1)(n-r)} + o(n).
\end{align*}
The last inequality is from the monotonicity and the eigen-spectrum of tridiagonal matrix.

Take $a = q^{H_q(\tau) n /} \log n$ and $V_q(n, r'') \leq a \leq V_q(n, r''+1)$. Thus $r'' = (\tau + o(1))n$. Then the reuslts follows from \eqref{eq:ball1}.

\end{proof}

\subsection{The $q$-ary first MRRW bound}

We now recover the established $q$-ary first MRRW bound. The following inequality is the specialization of \cite[Theorem 8.4]{Friedman2005Generalized} to the Hamming-graph covering. 

\begin{theorem}\label{Frithm}
\cite[Theorem 8.4]{Friedman2005Generalized}
$$
\lambda_2\left(\GCp\right) \geq \operatorname{FK}_{\mathcal{H}_q}(a)-\frac{n(q-1) a}{q^k} .
$$
\end{theorem}

\begin{theorem}\label{thm:MRRW}
Let $C_n$ be nonzero $q$-ary linear codes with relative distance $\delta$ and rate $R$. For $0\leq\delta\leq Q$,
\begin{equation}\label{eq:MRRW}
 R\le H_q \left(
 \frac{q-1-(q-2)\delta-2\sqrt{(q-1)\delta(1-\delta)}}q
 \right).
\end{equation}
\end{theorem}
\begin{proof}
Define $F_q(R) = n((q-2)R+2\sqrt{(q-1)R(1-R)})$. Take $a = q^{Rn}/ \log n$ in Theorem \ref{Frithm}, we have
$$
\lambda_2\left(\GCp\right) \geq \operatorname{FK}_{\mathcal{H}_q}(q^{Rn}/\log n)-o(n).
$$
\eqref{eq:ball2} implies
\[
 \frac{\lambda_2(\GCp)}{n}
 \ge F_q(H_q^{-1}(R))-o(1).
\]
Therefore,
\begin{equation}\label{eq:preinverse}
 q-1-q\delta\ge F_q(H_q^{-1}(R)).
\end{equation}

One can check that $F_q$ is strictly increasing on $[0,Q]$, and
\begin{equation}
  F_q^{-1}(q-1-q\delta)=
 \frac{\bigl(\sqrt{(q-1)(1-\delta)}-\sqrt\delta\bigr)^2}{q}.
\end{equation}

\end{proof}

\section{The Improvement of GV bounds}

Let $r\geq d-1$. The quotient Gilbert graph $\mathcal G_{C,r}$ has the cosets of $C$ as vertices. Two distinct cosets are adjacent when their distance is at most $r$. It is regular graph and its degree is $|B_C(0,r)|-1$. A union of cosets indexed by an independent set has minimum distance at least $d$, but is not necessarily linear. Therefore,

\begin{proposition} 
$$
A_q(n,d)\geq |C| \left\lceil \frac{|V(\mathcal{G}_{C,d-1})|}{|B_C(0,d-1)|} \right\rceil \geq \frac{q^n}{|B_C(0,d-1)|}.
$$
\end{proposition}

For $C=\{0\}$, it recovers the classical GV bound.  For a general auxiliary code, $|B_C(0,d-1)\leq V_q(n,d-1)$, so this  lower bound is at least as large.

For $x\in \Ftn$, denote $m_C(x,t) = |C\cap B_q(x,t)| = |\{c\in C\mid d(x,c) \leq t \}|$, i.e., the number of codewords in $C$ which has distance at most $t$ from $x$. 

\begin{lemma} 
$$
|B_C(0,r)| = \sum_{0\leq wt(x)\leq r} \frac{1}{m_C(x,r)}.
$$
\end{lemma}

\begin{proof}
Partition $B_q(0,r)$ according to the cosets of $C$. For each coset $Q$ intersecting $B_q(0,r)$, let $S_Q=Q\cap B_q(0,r)$. For any $x\in S_Q$, we have
\[
m_C(x,r)
=|C\cap B_q(x,r)|
=|(x+C)\cap B_q(0,r)|
=|S_Q|,
\]
where the second equality follows from translation invariance
of the Hamming distance and the fact that $-C=C$.

Consequently, $\sum_{x\in S_Q}\frac{1}{m_C(x,r)} =\frac{|S_Q|}{|S_Q|} =1$. Summing over all cosets $Q$ intersecting $B_q(0,r)$ yields the result.
\end{proof}

\begin{example} 
Assume $C=\{0,v\}$, where $v$ is a vector in $\Ftn$ with $\wt(v) = d$. Then $|B_C(0,r)| = V_2(n,r) - \frac{1}{2}D_2(n,d,r)$, where $D_2(n,d,r) = |B_2(0,r)\cap B_2(v,r)|$ is the size of two balls in $\Ftn$ with radius $t$ and distance $d$. An explicit binary intersection formula is
$$
 D_2(n,d,r)=\sum_{j=0}^r\ \sum_{i=0}^d
 \mathbf1\{d+j-2i\le r\}\binom{d}{i}\binom{n-d}{j-i}.
$$
Therefore, for any $r\geq d-1$,
$$
A_2(n,d)\geq 2 \left\lceil\frac{2^{n-1}}{V_2(n,r) - D_2(n,d,r)/2} \right\rceil.
$$ 
\end{example}

\begin{example} 
Let $v = (1^s 0^{n-s}) \in\Fqn$ and $C=\text{span}\{v\}$.  For $y\in\F_q^s$, define
$$
L(y,k) = \left|\{ a\in\F_q \mid |\{i:y_i=a\}|\geq k \}\right|
$$
and 
\begin{align*}
D(t,i)=  & |\{x = (y,z)\in\F_q^s\times\F_q^{n-s} \mid \wt(x) \leq t, \\
&L(y,s-t+\wt(z)) = i \}|.
\end{align*}

Thus the number of $x\in B_C(0,d-1)$ with $m_C(x,d-1) =i$ is exactly $D(d-1,i)$. Consequently, for every $s\ge d$,
\[
A_q(n,d)\geq q\left\lceil
\frac{q^{n-1}}{\sum_{i=1}^qD(d-1,i)/i}
\right\rceil.
\]
\end{example}

\begin{example}
Assume $n=sd$ with $d\geq2$ even, and let $C\subseteq\F_2^n$ be generated by the all-one vectors on $s$ disjoint blocks of size $d$. A coset in one block identifies a binary vector with its complement, so its minimum-weight enumerator is
\[
P_d(z)=\sum_{i=0}^{d/2-1}\binom di z^i
 +\frac12\binom d{d/2}z^{d/2}.
\]
Minimum coset weights add across blocks. Hence, writing $[z^j]$ for coefficient extraction,
\[
|B_C(0,d-1)|=\sum_{j=0}^{d-1}[z^j]P_d(z)^s,
\]
so
\[
A_2(n,d)\geq2^s\left\lceil\frac{2^{n-s}}{\sum_{j=0}^{d-1}[z^j]P_d(z)^s}\right\rceil.
\]
\end{example}

\begin{remark}
The bound can be further improved if $\Delta(\GCr)$ can be bounded from above for specific code $C$ with minimum distance $d$. 
For example, let $C\subseteq \Ftn$ be spanned by vectors of weight at most $w$, and suppose that $\bigcup_{\boldsymbol{c}\in C}\supp(\boldsymbol{c})=[n]$. Then, in \cite{shangguan2026improved}, the authors proved that for every $0\le \lambda\le 1$,

\begin{equation*}
|B_C(0,r)| \leq \lambda^{-r} \left(\frac{(1+\lambda)^w}{1+\lambda^w}\right)^{n/w}.
\end{equation*}

Taking $r=d-1$ gives
$$
A_2(n,d) \geq \frac{2^n}{\lambda^{-d}\left(\frac{(1+\lambda)^w}{1+\lambda^w}\right)^{n/w}}.
$$
\end{remark}

Algorithm~\ref{alg:1} starts from a specified linear code $C_0$ with minimum distance at least $d$; the choice $C_0=\{0\}$ is allowed. It chooses an admissible extension minimizing the number of occupied cosets. For $C_v=\operatorname{span}(C\cup\{v\})$,
\[
m_{C_v}(x,d-1)=\sum_{a\in\F_q}m_C(x+av,d-1).
\]
The test $d(v,C)\geq d$ guarantees that every nonzero new coset has minimum weight at least $d$, since $C$ is linear. The procedure is an existence and optimization framework.

\begin{algorithm}
\caption{Coset construction of code}\label{alg:1}
\begin{algorithmic}[1]
\REQUIRE $C_0\subseteq\F_q^n$ with $d(C_0)\geq d$
\STATE $C\gets C_0$
\STATE $m_C(x,d-1)\gets|C\cap B_q(x,r)|$ for all $x\in\F_q^n$
\WHILE{$|B_C(0,d-1)|<q^n/|C|$}
\STATE $b\gets+\infty$
\FOR{$v\in\F_q^n$ with $d(v,C)\geq d$}
\STATE $C_v\gets\operatorname{span}(C\cup\{v\})$
\STATE $m_{C_v}(x,d-1)\gets\sum_{a\in\F_q}m_C(x+av,d-1)$
\STATE $b_v\gets\sum_{x\in B_q(0,d-1)}1/m_{C_v}(x,d-1)$
\IF{$b_v<b$}
\STATE $b\gets b_v$; $u\gets v$
\ENDIF
\ENDFOR
\STATE $C\gets C_u$; $m_C(\cdot,d-1)\gets m_{C_u}(\cdot,d-1)$
\ENDWHILE
\RETURN $C$
\end{algorithmic}
\end{algorithm}

The next estimate quantifies the decrease that can be guaranteed at each step.

\begin{theorem}\label{thm:contraction}
If $B_C(0,d-1)\ne V(\mathcal{G}_{C,d-1})$, there is an admissible one-dimensional extension $C'$ with $d(C')\geq d$ such that
\begin{equation}\label{eq:contraction}
 |B_{C'}(0,d-1)|\leq |B_C(0,d-1)|-\kappa_q(d)
\end{equation}
\[
\kappa_q(d)=(q-1)\bigl(q^{d-1}-(q-1)^{d-1}\bigr).
\]
\end{theorem}

\begin{proof}
Since the radius of $\GC$ is at least $d$, a shortest path from zero to a sufficiently distant vertex contains
a vertex at distance exactly $d$. Choose a coset leader $v$ with $\wt(v)=d(v,C)=d$ and set $C'=\operatorname{span}(C\cup\{v\})$. Put $T=\supp(v)$ and $E_T=\{x:\supp(x)\subseteq T\}\cong\F_q^d$. We claim $C\cap E_T=\{0\}$. Otherwise, if $c\neq 0$ was supported in $T$, choose a nonzero scalar $a$ cancelling one nonzero coordinate of $v+ac$. No nonzero coordinate is created outside $T$, so $\wt(v+ac)\le d-1$, contradicting the leader property. Hence different vectors in $E_T$ are in the different cosets of $C$.

All $q^d-(q-1)^d$ vectors of $E_T$ with at least one zero coordinate have weight at most $d-1$, and therefore give that many distinct vertices of $\mathcal{G}_{C,d-1}$. For $\mathcal{G}_{C',d-1}$, $E_T$ is divided into $q^{d-1}$ cosets according to $v$. The number of neighbourhood loss is then at least
$q^d-(q-1)^d-q^{d-1}=\kappa_q(d)$ vertices.
\end{proof}

In particular, $\kappa_2(d)=2^{d-1}-1$. An extension minimizing the new occupied-coset count has at least this decrease.

\begin{theorem}\label{cor:linear}
Let $C_0$ have dimension $k_0$ and minimum distance $d$. Algorithm~\ref{alg:1} terminates after $\ell\geq0$ extensions with a linear code of dimension $k_0+\ell$ and minimum distance at least $d$, where
\begin{equation}\label{eq:implicit}
 q^{n-k_0-\ell}+\ell\kappa_q(d)\le |B_{C_0}(0,d-1)|.
\end{equation}
If $\ell_0$ is the smallest positive integer satisfying \eqref{eq:implicit}, then
\[
 A_q^{\rm lin}(n,d)\ge q^{k_0+\ell_0}.
\]
\end{theorem}

\begin{proof}
After $\ell$ extensions, Theorem~\ref{thm:contraction} bounds the occupied count by $|B_{C_0}(0,d-1)|-\ell\kappa_q(d)$. At termination the count equals the quotient size $q^{n-k_0-\ell}$, proving the results.
\end{proof}

\begin{remark}
For $w=\lfloor r/2 \rfloor$, note that $B_C(0,w)$ is a clique in $\GCr$. The size is $\sum_{1\leq \wt(x)\leq w} \frac{1}{m_C(x,d-1)} \geq V_q(n,w)/A_q'(n,d,w)$, where $A_q'(n,d,w)$ be the largest size of a distance-$d$ code contained in $B_q(0,w)$. Here $m_C(x,d-1) \leq A_q'(n,d,w)$ since $m_C(x,d-1)$ is the number of codewords in $C$ which has weight between $1$ and $w$. It follows that 
$$
A_q(n,d)\leq \frac{q^n A'_q(n,d,w)}{V_q(n,w)},
$$
which is the celebrated Elias--Bassalygo inequality.
\end{remark}

\section{Conclusion}

In this paper, we use coset graphs to study several parameters of linear codes and their duals. The spectrum of a coset graph encodes codeword weights, while its metric and expansion properties reflect quantities such as minimum distance, dual distance, and covering radius.

Using closed walks, non-backtracking walks, exponential kernels, Cheeger inequalities and random walks, we derive several bounds and relations among coding parameters. We also extend the framework to $q$-ary codes and obtain asymptotic rate-versus-distance bounds. In addition, a quotient version of the Gilbert graph incorporates overlaps among Hamming balls and leads to refinements of the GV lower bound.

Although some of our results do not improve the existing bounds, they provide a new perspective to illustrate how different graph-theoretic properties can be used to study coding-theoretic upper and lower bounds within a common setting, and may offer further improvements when additional graphic structure is available.

\end{document}